\documentstyle[12pt]{article}

\def\bp{{\bf p}}
\def\bk{{\bf k}}
\def\bq{{\bf q}}

\def\err{\end{array}}

\begin{document}

\title{Propagator in the instanton background and instanton-induced
processes in scalar model}

\author{ Yu.A. Kubyshin \\
{\small \em Institute for Nuclear Physics, Moscow State University}  \\
{\small \em 117899 Moscow, Russia \vspace{0.2cm} }\\ \vspace{0.2cm}
and P.G. Tinyakov \\
{\small \em Institute for Nuclear Research of the Russian Academy of 
Sciences}\\
{\small \em 60th October Anniversary prospect, 7a 
117312 Moscow, Russia } 
} 

\date{}

\maketitle

\begin{abstract} 
{\small 
The cross section of the multiparticle scattering processes in the
non-per\-tur\-ba\-tive sector of the scalar $(-\lambda) \phi^{4}$
model is studied within the semiclassical approximation. For this
purpose the exact formula for the residue of the propagator in the
instanton background is derived. The exponent of the cross section is
calculated at small energies both analytically and numerically in the
leading and next-to-leading orders in energy. The results are in
agreement with the numerical result of Ref. \cite{KT}.}
\end{abstract}

\section{Introduction}

In the present paper we study non-perturbative contributions to
scattering processes in the simple scalar model. The motivation is
that analogous processes play an important role in more realistic
theories. One class of such processes includes winding number
transitions between topologically inequivalent vacua in non-abelian
gauge theories and sigma models. In a number of papers (for a review
see, e.g., Refs.\cite{Ti1}) instanton transitions induced by particle
collisions in the Electroweak Theory were studied.  In QCD,
instanton-like processes in the deep inelastic scattering and their
possible experimental detection are under intensive investigation now
\cite{Ri2}.

The decay of metastable (false) vacuum \cite{VKO} is another
well-known example where tunneling plays an important role. It occurs
in scalar theories.  The solution which interpolates between initial
and final state is the bounce configuration \cite{Coleman}. The bounce
solution has one turning point; the field configuration at the turning
point represents the final state after the false vacuum decay has
taken place.  At low energies the amplitude of the false vacuum decay
is proportional to $\exp(-S_B/2)$, where $S_B/2$ is the Euclidean
action of the bounce configuration calculated up to the turning point
($S_B$ is the action of the whole bounce solution).

Finally, there are so-called shadow processes first considered in
Refs. \cite{Vo1,Hsu}. These are the processes we concentrate on in
this paper. They arise in models with false vacuum decay and
correspond to transitions between initial and final states, both lying
in the false vacuum, through the intermediate state containing the
bubble of the true vacuum. At low energies, such transitions are dominated
by the same bounce solution, the difference with the false vacuum
decay being that the final state is not the one corresponding to the
turning point, but to asymptotically large Euclidean times.  Thus,
amplitudes of such processes are proportional to $\exp(-S_B)$. Clearly
these processes are unphysical since their probabilities are much
smaller than the probability of the false vacuum decay. Nevertheless,
formally these processes are fully analogous to the instanton-like
transitions in gauge theories and, because of relative simplicity of
scalar models, can serve as a good laboratory for testing methods of
calculation of probabilities of instanton-like transitions.

The main method for studying instanton-like transitions at non-zero
energies is the generalization of the semiclassical approximation. It
can be shown that instanton contribution into the total
cross section of the inclusive process $N \rightarrow $ {\it any} with
the initial energy $E$ is given by the semiclassical expression
\cite{RT,Ti2}
\begin{equation}
\sigma_{N} (E) =  \sum_{M} \sigma_{N \rightarrow M}(E)  
\sim e^{\frac{1}{\lambda} F(\epsilon, \nu) + {\cal O}(\lambda^{0})}, 
\label{sigma-gen}
\end{equation}
where $\lambda$ is the coupling constant, $\epsilon = E/E_{sph}$, $\nu
= N/N_{sph}$, and $E_{sph}$, $N_{sph}$ are the sphaleron energy and
the sphaleron number of particles, respectively\footnote{The number
of particles in the sphaleron is defined as the number of particles to
which the sphaleron decays if slightly perturbed \cite{HK}.  In the
case of false vacuum decay or shadow processes the analog of the
sphaleron is a critical bubble \cite{VKO}.}.  In the limit of very low
energy of initial particles the exponent in Eq. (\ref{sigma-gen}) is
equal to
\begin{equation}
\frac{1}{\lambda} F(0, 0) = -2 S_{inst}   \label{F0}  
\end{equation}
and the cross section is suppressed by the inverse power of 
the small coupling constant.
However, at parametrically high energy $E \sim E_{sph} \sim
1/\lambda$, the cross section depends exponentially on $E$.  This
leads to the reduction of the suppression factor at energies of the
order of the sphaleron energy, so that instanton induced processes
may become significant. The main application of this approach is based
on the conjecture of Refs. \cite{RT,Ti2} that in the limit $\nu\to 0$ 
the function $F(\epsilon, \nu)$ reproduces the exponent in the 
instanton contribution into the total cross section of the 
{\em two-particle} scattering and thus allows to estimate the latter. 

The induced false vacuum decay in the $(- \lambda) \phi^{4}$ model was
studied in Ref. \cite{KT}. There the function $F(\epsilon, \nu)$ was
computed numerically in the range of parameters $ 0.4 \leq \epsilon
\leq 3.5$ and $0.25 \leq \nu \leq 1.0$. The calculation was based on a
numerical realization of the formalism of Refs. \cite{RT} - \cite{RST}
involving solution of a certain classical boundary value problem.  In
the present paper we study the case of shadow processes. We limit
ourselves to perturbation theory around the instanton. Although such
perturbation theory works only at $E\ll E_{sph}$ and does not provide
an explicit procedure for calculating the function $F(\epsilon, \nu)$
in a closed form, it still remains one of the main theoretical tools
for studying instanton-like transitions.

Similar to the conventional perturbation theory, the key role in the
perturbative expansion around the instanton is played by the
propagator in the instanton background. The scattering amplitudes, as
well as the contributions to the function $F(\epsilon, \nu)$, are
expressed through the on-mass-shell residues of the propagator and
vertices in the external instanton field. While in most models
possessing instantons (in particular, in the Electroweak Theory) the
propagator in the instanton background is not known explicitly, this
is not the case in the model we consider here. Making use of this
advantage, we first compute exactly the residue of the propagator in
the instanton background found in Ref. \cite{Ku}. We then calculate
the leading and next-to-leading corrections to the zero energy value
(\ref{F0}) of the function $F(\epsilon, \nu)$. This approximation is
valid for low energies, and besides being useful on its own, enables
one to obtain the behavior of the exponential factor in the range of
parameters which was not covered by the numerical computations in
Ref. \cite{KT}. In the low energy limit analytical expressions for
$F(\epsilon, \nu)$ can be obtained. The latter allows us to check (in
this approximation) the validity of the conjecture of
Refs. \cite{RT,Ti2} about the relation between two-particle and
multiparticle cross sections. This conjecture essentially relies on
the regularity of the function $F(\epsilon, \nu)$ in the limit $\nu
\rightarrow 0$, and we show that indeed all singular terms, appearing
at the intermediate stage of the calculation, cancel out in the final
expression.
 
This paper is organized as follows. In Sect. 2 we describe the model
and calculate the residue of the propagator in the instanton
background. In Sect. 3 we calculate the function $F(\epsilon, \nu)$ in
the leading and next-to-leading approximations. The results are
compared to numerical results of Ref.~\cite{KT} in the region where
the latter can be translated to the case of shadow processes. Analysis
of the expression for the suppression factor in the limit of small
$\epsilon$ and $\nu$ will also be given. Sect. 4 contains discussion
and concluding remarks.

\section{Propagator in the instanton background and its \\
residue}

Consider the model of one real scalar field with the action 
\begin{equation}
 S = \int d^{4}x \left( \frac{1}{2} \partial_{\mu} \phi 
     \partial^{\mu} \phi - \frac{1}{2} m^{2} \phi^{2} + 
     \frac{\lambda}{4!} \phi^{4} \right)     \label{action}
\end{equation}
in the four-dimensional Minkowski space-time, where $\lambda > 0$. For 
$m \neq 0$ the potential has the meta-stable minimum at $\phi = 0$ and 
is unbounded from below when $|\phi| \rightarrow \infty$. 

In the case $m=0$ the state $\phi = 0$ is stable classically but is
unstable with respect to quantum fluctuations. Its decay is described
by the instanton solutions \cite{Fu,Lip},
\begin{equation}
   \phi_{inst} (x) =  \frac{4 \sqrt{3}}{\sqrt{\lambda}} 
      \frac{\rho}{(x-x_{0})^{2}+\rho^{2}}      \label{inst}
\end{equation}
parameterized by the size $\rho$ and position $x_{0}$. Due to conformal
invariance of the theory (\ref{action}) in the massless case, the size
$\rho$ can be arbitrary and the action of the instanton configuration
\[
S_{inst} = \frac{16 \pi^{2}}{\lambda}   
\]
does not depend on it. For the perturbative calculations of the
function $F(\epsilon, \nu)$ the on-mass-shell residue of the instanton
solution will be needed. By definition it equals
\[
R_{inst}({\bf p}) = \frac{1}{(2 \pi^{2})^{3/2} \sqrt{2 w_{\bf p}}} 
\lim_{p^{2} \rightarrow -m^{2}} (p^{2}+m^{2}) 
\tilde{\phi}_{inst}(p), 
\]
where $\tilde{\phi}_{inst}(p)$ is the Fourier transform of the 
instanton, 
\[ 
 \tilde{\phi}_{inst}(p) = \int d^{4}x e^{ipx} \phi_{inst}(x).   
\]
{}From explicit expression (\ref{inst}) one obtains  
\begin{equation}
  \tilde{\phi}_{inst}(p) = \frac{16 \sqrt{3} \pi^{2}}{\sqrt{\lambda}} 
  \frac{\rho^{2}}{|p|} K_{1}(\rho |p|), \label{inst-FT}
\end{equation}
where $K_{1}(z)$ is the modified Bessel function, and  
\begin{equation}
  R_{inst}({\bf p}) = \frac{1}{\sqrt{\lambda}} \frac{16 \sqrt{3} 
  \pi^{2} \rho}{(2 \pi^{2})^{3/2} \sqrt{2 w_{\bf p}}}.  \label{R-inst}
\end{equation}

For $m \neq 0$ regular Euclidean solutions with finite action do 
not exist. The decay of the meta-stable state $\phi = 0$ is dominated by 
approximate solutions known as constrained instantons \cite{Aff}. 
They minimize the Euclidean action of the theory under the 
constraint that the size of the configuration is $\rho$. 
The conformal invariance is broken in the massive case and the action 
depends on $\rho$ and on the constraint. For $\rho^{2} m^{2} \ll 1$ 
the constraint instanton configuration coincides with the massless 
instanton (\ref{inst}) at $|x| \ll \rho$ and decreases as 
\[
 \phi_{inst} (x) \sim \frac{2 \sqrt{6 \pi}}{\sqrt{\lambda}} 
   \frac{\rho \sqrt{m}}{|x|^{3/2}} e^{-m|x|} 
\]
for $|x| \ge m^{-1}$. In this case the constrainet-independent part of
the instanton action has the form \cite{KT}
\begin{equation}
 S_{inst} = \frac{16 \pi^{2}}{\lambda} - \frac{24 \pi^{2}}{\lambda}
  (\rho m)^{2} \left[ \ln \frac{(\rho m)^{2}}{4} + 2 C_{E} + 1 \right] 
  + {\cal O} \left( (\rho m)^{4} \right),     \label{inst-action1}
\end{equation}
where $C_{E} = 0.577 \ldots $ is the Euler constant (notice the
difference in the normalization of the coupling constant in
Eq. (\ref{action}) and in Ref. \cite{KT}). The constraint dependence
appears only in the terms of the order ${\cal O} \left( (\rho m)^{4}
\right)$ and higher. Similarly, other quantities of interest, in
particular, the residues of the instanton field and propagator in the
instanton background, are series in the parameter $m^2\rho^2$. To our
approximation only the leading term is important, and thus we can
calculate residues in the massless theory.

In the massive case the height of the barrier separating the vacuum
$\phi = 0$ and the instability region is finite and is determined by a
static sphaleron configuration found numerically in \cite{KT}. From
this solution one gets
\[
E_{sph} = \kappa \frac{m}{\lambda}, \; \; \; 
N_{sph} = \frac{\delta}{\lambda},    
\]
where the numerical factors $\kappa$ and $\delta$ are equal to 
$\kappa=113.4$, $\delta=63$. 

Now let us turn to the discussion of the propagator in the instanton
background and to calculation of its residue. The expression for the
instanton propagator in the massless case was obtained in Ref.
\cite{Ku} and used there for the computation of the first correction
to the asymptotic formula for the large-order coefficients of the
perturbation theory expansion \cite{Lip}, \cite{asymp}.

The equation for the propagator $G(x,y)$ in the background of the
instanton (\ref{inst}) reads
\begin{equation}
\Bigl\{ - \frac{\partial}{\partial x^{\mu}} 
\frac{\partial}{\partial x^{\mu}} - \frac{\lambda}{2} 
\phi^2_{inst}(x)\Bigr\} G(x,y) = 
\delta^4(x-y) - \sum_{A=1}^{5} {\rho^2 \psi_{A}(x) \psi_{A}(y)
\over (x^2+\rho^2)^2}, \label{prop-eq}
\end{equation}
{\tolerance=500
where $\psi_{A}(x)$ are zero modes of quadratic action describing
fluctuations around the instanton.

}
In calculation of the propagator in \cite{Ku} the $O(5)$-symmetry of
the theory (\ref{action}) with $m=0$ is essential.  After making the
stereographic projection of the four-dimensional Euclidean space
$E^{4}$ onto the sphere $S^{4}$, embedded into the five-dimensional
Euclidean space, Eq. (\ref{prop-eq}) simplifies drastically and becomes
the equation for the free propagator on $S^{4}$. Then, representing the
propagator in terms of the spherical harmonics and using the summation
formulas, one derives the explicit expression for $G(x,y)$. Perhaps it
is not surprising that tools used for the computation of large-order
coefficients of perturbation theory find also their application in the
computation of the non-perturbative cross-section at high energies.
Deep connections between these two problems have been observed in
Refs. \cite{ZaMa}.

The propagator in the instanton background in terms of 
the coordinates in $E^{4}$ is equal to
\begin{eqnarray}
G(x,y) & = & \frac{1}{2\pi^{2}} \frac{\rho^{2}}{(\rho^{2} + x^{2}) 
(\rho^{2} + y^{2})} \{ \frac{ 3 t^{2}(x,y) -1}{2} \frac{1}{1-t(x,y)} 
\nonumber \\ 
       & - & 3 t(x,y) \ln \frac{1-t(x,y)}{2} - \frac{ 41}{10} t(x,y) - 
\frac{3}{2} \}, 
\label{propag}
\end{eqnarray}
where as before $\rho$ is the size of the instanton.  The function
$t(x_{1},x_{2})$ is defined through the geodesic distance $d(\xi
_{1},\xi_{2})$ between the points $\xi_{1}$ and $\xi_{2}$ on the
sphere $S^{4}$ which correspond to the points $x_{1}$ and $x_{2}$ in
the four-dimensional Euclidean space:
\[ 
t(x_{1},x_{2}) = \cos (d(\xi _{1},\xi_{2})). 
\]
In terms of the coordinates in $E^{4}$ the function $t(x_{1},x_{2})$
equals to
\begin{equation}
 t(x_{1},x_{2}) = 1 - \frac{2 \rho^{2}(x_{1} - x_{2})^{2}} 
{(\rho^{2} + x_{1}^{2})(\rho^{2} + x_{1}^{2})}.
\label{t}
\end{equation}
The limit when points $x_{1}$ and $x_{2}$ coincide corresponds to
$t(x_{1},x_{2}) \rightarrow 1$. The propagator (\ref{propag}) 
is singular in this limit, as it should be.  For $x \rightarrow y$ 
its leading singularity is given by
\[ 
G(x,y) \sim \frac{1}{2\pi^{2}} \frac{\rho^{2}}{(\rho^{2} + x^{2}) 
(\rho^{2} + y^{2})}  \frac{1}{1-t(x,y)} + \mbox{weaker 
singularities}. 
\] 
Using Eq. (\ref{t}) the leading singularity can be transformed
into the expression which coincides with the free propagator of a
scalar field:
\[ 
G_{0}(x,y) = \frac{1}{4 \pi^{2}} \frac{1}{(x-y)^{2}}.
\]
This result is quite clear. Indeed, it can be shown that the
diagram, corresponding to the propagator in the instanton background,
can be represented as an infinite sum of diagrams $D_{n}$. Each
$D_{n}$ consists of a line, representing the free propagator, with $n$
insertions of two instanton fields each, so that $D_{n}$ has two
external lines of the quantum free field and $2n$ external lines of
the classical instanton field.  Analytically the relation between the
full instanton propagator and the sum of the free propagators with the
insertions is written as follows:
\begin{eqnarray}
G(x_{1},x_{2}) & = & G_{0}(x_{1},x_{2}) + 
\frac{\lambda}{2} \int dy G_{0}(x_{1},y) \phi_{inst}^{2}(y) 
G_{0}(y,x_{2}) + \ldots     \nonumber  \\
  & = & \frac{1}{4 \pi^{2}} \frac{1}{(x_{1}-x_{2})^{2}} - 
  \frac{3}{2 \pi^{2}} \frac{\rho^{2}}{(\rho^{2} + x_{1}^{2}) 
  (\rho^{2} + x_{2}^{2}) - \rho^{2} (x_{1} - x_{2})^{2}} \nonumber \\
  & \times & \ln \frac{\rho^{2} (x_{1} - x_{2})^{2}}
  {(\rho^{2} + x_{1}^{2})(\rho^{2} + x_{2}^{2})} + \ldots \nonumber
\end{eqnarray}
The first term, which is just the free propagator, corresponds to the
diagram $D_{0}$ and is the most singular one.  The subleading
singularity, given by the second term, corresponds to $D_{1}$ and is
of the logarithmic type, in agreement with the complete expression
(\ref{propag}).

As it was already said in the Introduction, for the calculation of the
exponential factor in Eq. (\ref{sigma-gen}) in the next-to-leading
approximation one needs the double on-mass-shell residue $R(p_{1},
p_{2})$ of the propagator in the instanton background.  This function
is defined as
\begin{equation}
R(p_{1},p_{2}) = \lim_{p_{1,2}^{2} \rightarrow -m_{1,2}^{2}} 
(p_{1}^{2} + m_{1}^{2}) (p_{2}^{2} + m_{2}^{2}) \tilde{G}(p_{1},p_{2}),
\label{res-gen}
\end{equation}
where $\tilde{G}(p_{1},p_{2})$ is the standard Fourier transform,
\[
\tilde{G}(p_{1},p_{2}) = \int d^{4}x d^{4}y G(x,y) 
\exp (ip_{1}x + ip_{2}y).
\]
In the massless approximation the residue $R(p_{1},p_{2})$ is actually
a function of $\rho^{2}s$ only, where the variable $s$ is defined by
$s \equiv s(p_{1},p_{2}) = (p_{1} + p_{2})^{2} = 2 (p_{1}p_{2})$.

Calculation of $R(p_{1},p_{2})$ is rather tedious although
straightforward procedure, and we do not give the details
here. However, we would like to discuss some general features of the
computation before presenting the answer.

The terms in the expression (\ref{propag}) give contributions to the
residue which can be divided in four classes.

1) The first term in the curly brackets in eq. (\ref{propag}) gives
rise to the free propagator term as it was already explained.  Its
contribution to the Fourier transform $\tilde{G}(p_{1},p_{2})$ is
proportional to
\[ 
\frac{2 (2 \pi)^{4}}{p_{1}^{2}} \delta (p_{1} + p_{2}). 
\]
This describes free motion of the particle not interacting with the 
instanton and is irrelevant for our problem.

2) There are factorizable terms of the form $f_{1}(x) f_{2}(y)$, where
$f_{i}(x)$'s are proportional to expressions like
\begin{equation}
 \frac{x^{n}}{(\rho^{2} + x^{2})^{m}} \; \; \; \mbox{or} \; \; 
\; \frac{x^{n} \ln(\rho^{2} + x^{2})}{(\rho^{2} + x^{2})^{m}} 
\label{fterms}
\end{equation}
with some integer $n$ and $m$. Their contributions to the 
momentum-space propagator are of the form $\tilde{f}_{1}(p_{1}^{2}) 
\tilde{f}_{2}(p_{2}^{2})$. These obviously give $s$ - independent 
contributions to the residue (\ref{res-gen}).

3) The next group of terms are of the form $(xy) f_{1}(x) f_{2}(y)$
with $f_{i}$ given by (\ref{fterms}). Calculating the momentum-space
propagator we get
\begin{eqnarray}
 \int e^{ip_{1}x + ip_{2}y} (x,y) f_{1}(x) f_{2}(y) & = & - 
\frac{\partial}{\partial p^{1\mu}} 
\frac{\partial}{\partial p_{2}^{\mu}}
\int e^{ip_{1}x + ip_{2}y} f_{1}(x) f_{2}(y) \nonumber \\
= - \frac{\partial}{\partial p^{1\mu}} 
\frac{\partial}{\partial p_{2}^{\mu}}
\tilde{f}_{1}(p_{1}^{2}) 
\tilde{f}_{2}(p_{2}^{2}) & = & - 4 (p_{1},p_{2}) 
\tilde{f}'_{1}(p_{1}^{2}) \tilde{f}'_{2}(p_{2}^{2}).
\nonumber
\end{eqnarray}
This gives a contribution to the residue proportional to 
$s(p_{1},p_{2})$.

4) The last group consists of terms of the form $(xy) \ln
(x-y)^{2} f_{1}(x) f_{2}(y)$ and $\ln (x-y)^{2} f_{1}(x)
f_{2}(y)$. Carrying out the computations one can show that they lead
to terms proportional to $s \ln s$ and $\ln s$ respectively in the
expression for the function (\ref{res-gen}).

Finally, the expression for the residue of the momentum-space
propagator with both momenta on the mass shell is equal to
\begin{equation}
 R(s) = 16 \pi^{2} \left[ \frac{3}{4} s \ln \frac{s}{\sqrt{2}} + 
 \frac{3}{2} s \left( C_{E} - \frac{1}{15}\right) - \frac{3}{2} 
 \ln \frac{s}{\sqrt{2}} + 3 \left(C_{E} - \frac{43}{30}\right) \right].
\label{res}
\end{equation}
 
This result agrees with the asymptotic formula for large $s$ derived
in Ref. \cite{Vo2}:
\begin{equation}
R(s) = \lim_{p_{1,2}^{2} \rightarrow 0} p_{1}^{2} p_{2}^{2} 
 \frac{\tilde{\phi}_{inst}(p_{1})\tilde{\phi}_{inst}(p_{2})}
 {(\partial_{\mu} \phi_{inst}, r^{-2} 
\partial^{\mu} \phi_{inst})} (p_{1},p_{2}) \ln (p_{1},p_{2}).
\label{asymp}
\end{equation}
Indeed, using the expression (\ref{inst-FT}) for the Fourier transform
of the instanton solution and working out the formula (\ref{asymp})
one can see that it coincides with the leading asymptotic of the exact
expression (\ref{res}) for large $s$:
\[ 
R(s) \sim 12 \pi^{2} s \ln s. 
\]

\section{Multiparticle cross section}

The formula (\ref{sigma-gen}) for the multiparticle cross-section of 
shadow processes comes from the following expression derived in 
Ref. \cite{Ti2},
\begin{eqnarray}
\sigma_{N} (E) & \sim & \int d^{4}x d \rho d^{4}\xi d \theta 
\exp \left[ - 2 S_{inst}(\rho) + 
\frac{1}{\lambda} W^{(1)}(x_{0}, \rho, \xi, \theta) \right. \nonumber \\
& + & \left.
\frac{1}{\lambda} W^{(2)}(x_{0}, \rho, \xi, \theta) + \ldots \right], 
\label{sigma-W}
\end{eqnarray} 
where we integrate over the position $x$ and the size $\rho$ of the
instantons, as well as over auxiliary variables $\xi_{\mu}$ and
$\theta$.  We also indicated explicitly the dependence of the action
on the size of the instanton (see Eq.  (\ref{inst-action1})). The
terms $W^{(i)}$ account for fluctuations in the instanton background:
$W^{(1)}$ corresponds to leading diagrams without propagator
insertions, $W^{(2)}$ corresponds to diagrams with one internal
propagator in the instanton background, etc. Diagrams with loops do
not appear in the $(1/\lambda)$ order of the semi-classical
approximation, they contribute to ${\cal O} (\lambda^{0})$ terms in
(\ref{sigma-gen}). The integrals in (\ref{sigma-W}) are evaluated by
the saddle point method. Simple dimensional analysis shows that the
result (the function $F$ in (\ref{sigma-gen})), actually
depends on $\epsilon = E/E_{sph}$ and $\nu=N/N_{sph}$ only.

In the present paper we limit ourselves to the calculation of the first
two contributions $W^{(1)}$ and $W^{(2)}$. General expressions for
these functions, derived in \cite{Ti2}, are quite cumbersome and are
given in the Appendix. It can be seen that up to the next-to-leading
order the saddle point values of $x$, $\rho$, $\xi$ and $\theta$
are determined by the leading-order equations. These equations are
obtained by differentiation of the expression $S_{inst} +
W^{(1)}/\lambda$, where $S_{inst}$ and $W^{(1)}$ are given by the
expressions (\ref{inst-action1}) and (\ref{W1-1}), with respect to
$x_{0}$, $\rho$, $\xi$ and $\theta$. The physically relevant saddle
point has $x_i=0$, $\xi_i=0$ ($i=1,2,3$), while $x_{0}$, $\xi_{0}$ and
$\theta$ are purely imaginary. It is convenient to introduce the
following notations,
\begin{equation}
x_{0}= i\tau, \; \; \; \xi_{0} = i\chi, \; \; \; 
\theta = -i \ln \gamma.   \label{x-tau}
\end{equation} 

For general values of $\epsilon$ and $\nu$, the system of saddle point
equations is transcendental and cannot be solved analytically even in
the leading order. It simplifies considerably in the limit of small
$\nu$. To the leading order in $\nu$ the saddle point solution can be
written in the form
\begin{eqnarray} 
\tilde{\rho}^{2} & = & - \frac{\kappa}{192 \pi^{2} m^{2}} 
\frac{\epsilon}{\Phi'(\tilde{\tau})},    \label{saddle-r2} \\
\tilde{\gamma} & = & - \frac{1}{4} 
\left( \frac{ \delta \nu}{\kappa \epsilon} \right)^{3} \Phi' 
(\tilde{\tau}),       \label{saddle-g} \\
\tilde{\chi} & = & \tilde{\tau} + \frac{2}{m} 
\frac{\delta \nu}{\kappa \epsilon}, \label{saddle-x}
\end{eqnarray}
where  the function $\tilde{\tau}=\tilde{\tau}(\epsilon,\nu)$ 
is determined by the equation 
\[
\ln \left( - \frac{\kappa \epsilon C e}{192 \pi^{2} \Phi' (\tilde{\tau})} 
  \right) + 4 \left[ \Phi (\tilde{\tau}) - 
  \frac{\delta \nu}{\kappa \epsilon} 
  \Phi' (\tilde{\tau}) + {\cal O}(\nu^{2}) \right] = 0. 
\]  
Here
\[
\Phi(\tau) \equiv \frac{m}{\tau} K_{1}(m\tau),
\]
the prime denotes the derivative with respect to $m\tau$ and 
$C = - \ln 4 + 2 C_{E} + 1$. 

It is straightforward although lengthy calculation to substitute these
expressions into the exponent in Eq. (\ref{sigma-W}) and check that the
limit $\nu\to 0$ is smooth and singular terms appearing at the
intermediate stages of the calculation cancel in the final answer. We
will show this in the low energy limit. At $\epsilon\to 0$, the saddle
point equations simplify further. The solution for 
$\tilde{\tau} (\epsilon, \nu)$ is 
\[
   \tilde{\tau} = \tilde{\tau}_{0}(\epsilon) + 
   \delta \nu \frac{8}{\kappa m \epsilon \left( 8 - 
   3 (m \tilde{\tau}_{0}(\epsilon))^{2} \right)}, 
\]
where $\tilde{\tau}_{0}(\epsilon)$ is determined by the equation 
\[
\ln \left( \kappa \epsilon \frac{C e}{394 \pi^{2}}
  (m \tilde{\tau}_{0})^{3}\right) 
  = - \frac{4}{( m \tilde{\tau}_{0})^{2}}.   
\]
The last equation can be solved iteratively, and one gets
\[
  m \tilde{\tau}_{0}(\epsilon)  = \frac{2}{\sqrt{\ln \frac{1}{\epsilon}}}
  + \frac{\ln \ln \frac{1}{\epsilon}}
  {\left( \ln \frac{1}{\epsilon} \right)^{3/2}} + \ldots
\]
Using relations (\ref{saddle-r2}) - (\ref{saddle-x}) one finds in 
the leading order 
\begin{eqnarray}
(m \tilde{\rho})^{2} & = & \frac{1}{48 \pi^{2}}
\frac{\kappa \epsilon}{\left( \ln \frac{1}{\epsilon} \right)^{3/2}}, 
\nonumber  \\
\tilde{\gamma} & = & \left( \frac{\delta \nu}{\kappa \epsilon} \right)^{3} 
\left( \ln \frac{1}{\epsilon} \right)^{3/2}, \nonumber \\
\tilde{\chi} - \tilde{\tau} & = & \frac{2}{m} 
\frac{\delta \nu}{\kappa \epsilon}.   \nonumber
\end{eqnarray}

To the leading order in energy and zeroth order in $\nu$ 
the function $F(\epsilon, \nu)$ can be written as 
\[
F(\epsilon, \nu) = -32 \pi^{2} + F^{(1)}(\epsilon, \nu) + 
  F^{(2)}(\epsilon, \nu) + \ldots, 
\]
where 
\begin{equation}
F^{(1)}(\epsilon, \nu) = 2 \frac{\kappa \epsilon}
{\sqrt{ \ln \frac{1}{\epsilon}}} \left[ 1 + {\cal O} 
\left( \frac{\ln \ln \frac{1}{\epsilon}}{\ln \frac{1}{\epsilon}} 
\right) \right] + {\cal O}(\nu)    \label{F1}
\end{equation}
and
\[
F^{(2)}(\epsilon, \nu) = F^{(2)}_{(f-f)}(\epsilon, \nu)+ 
F^{(2)}_{(i-f)}(\epsilon, \nu) + F^{(2)}_{(i-i)}(\epsilon, \nu) 
\]
is the sum of contributions coming from final-final, initial-final and 
initial-initial particle interactions,   
\begin{eqnarray}
F^{(2)}_{(f-f)}(\epsilon, \nu) & = & 
- \frac{1}{128 \pi^{2}} (\kappa \epsilon m \tilde{\tau}_{0})^{2}
\left[ \frac{58}{15} + \ln \left( 
\frac{\kappa \epsilon m \tilde{\tau}_{0}}{384 \pi^{2}} \right) \right], 
\label{F2-ff} \\
F^{(2)}_{(i-f)}(\epsilon, \nu) & = & 
\frac{6}{(192 \pi^{2})^{2}} (\kappa \epsilon m \tilde{\tau}_{0})^{3}
\left[ \frac{71}{30} + \ln \left( 
\frac{(\kappa \epsilon m \tilde{\tau}_{0})^{2}}{768 \pi^{2}} 
\right) + \ln \frac{1}{\delta \nu} \right], 
\label{F2-if} \\  
F^{(2)}_{(i-i)}(\epsilon, \nu) & = & 
- \frac{3}{(192 \pi^{2})^{2}} (\kappa \epsilon m \tilde{\tau}_{0})^{3}
\left[ \frac{71}{30} + \ln \left( 
\frac{(\kappa \epsilon m \tilde{\tau}_{0})^{3}}{1536 \pi^{2}} 
\right) + 2 \ln \frac{1}{\delta \nu} \right]. 
\label{F2-ii}     
\end{eqnarray} 
We see that $F^{(2)}_{(i-i)}$ and $F^{(2)}_{(i-i)}$ contain terms
$\ln (1/\nu)$ which are singular in the limit $\nu \rightarrow
0$. However, when (\ref{F2-ff}) - (\ref{F2-ii}) are summed together
the singular terms cancel each other. Finally, we get
\begin{equation}
F^{(2)}(\epsilon, \nu) = \frac{(\kappa \epsilon)^{2}}{32 \pi^{2}}
 \left( 1 + \frac{1}{2} 
 \frac{\ln \ln \frac{1}{\epsilon}}{\ln \frac{1}{\epsilon}} \right)
 + \ldots
\label{F2}
\end{equation} 
{}From Eqs. (\ref{F1}), (\ref{W2}) we see that 
our approximation is valid as long as 
\begin{equation}
\frac{\ln \ln \frac{1}{\epsilon}}{\ln \frac{1}{\epsilon}} \ll 1.
\label{e}
\end{equation}

At higher energies the condition (\ref{e}) breaks down, and the
calculation has to be done numerically.  We have performed this
calculation for the values of $\nu$ not subject to the condition
$\nu\ll 1$. For the periodic instanton case (the values of $\epsilon$
and $\nu$ such that $\tilde{\gamma}=1$ and $\tilde{\chi} =
2\tilde{\tau}$) the results can be compared to those obtained in
Ref.\cite{KT} and show good agreement.  Moreover, the next-to-leading
order correction $F^{(2)}(\epsilon, \nu)$ improves systematically the
agreement as compared to the previous order expression $(- 32 \pi^{2}) +
F^{(1)}(\epsilon, \nu)$.

\section{Discussion and conclusions}

In the present paper we have analyzed the multiparticle cross section
of the shadow processes induced by instanton transitions in the simple
scalar model (\ref{action}). We have obtained the exact analytical
expression for the on-shell residue of the propagator of quantum
fluctuations in the instanton background. This allowed us to calculate
the suppression factor in the next-to-leading order.
The range of validity of this approximation can be estimated by
comparing the results with numerical computations of the complete
function $F(\epsilon, \nu)$ in Ref. \cite{KT} in the range where the
latter can be translated to the case of shadow processes (i.e., for
periodic instantons).  The comparison shows that the perturbative
results do not differ significantly from the exact ones for $\epsilon
\leq 0.25$ and $\nu \leq 0.2$. This range can be taken as the region
of validity of our perturbative calculation.

For very small energies, namely when the condition (\ref{e}) holds, we
have obtained the analytical expressions for the suppression factor and
values of the saddle point parameters $\rho$, $\chi$, $\tau$ and
$\theta$. This enables us to check the validity of the approximation
used. For instance, the constraint-dependent contributions to the
action (and thus to the function $F(\epsilon, \nu)$) are of the
order $(\rho m)^4 \sim \epsilon^2/\ln^3 (1/\epsilon)$ and are
subleading as compared to the terms retained in the function
$F^{(2)}(\epsilon, \nu)$, Eq. (\ref{F2}). Also, this calculation
allowed us to check explicitly the cancelation of the terms singular
in the limit $\nu\to 0$ in the propagator contribution
$F^{(2)}(\epsilon, \nu)$.

\noindent{\large \bf Acknowledgments}

{\tolerance=500
We would like to thank A. Ringwald and V. Rubakov for
discussions and valuable comments.  Y.K. acknowledges financial
support from the Russian Foundation for Basic Research (grant
98-02-16769-a) and grant CERN/S/FAE/1177/97. The work of P.T. is
supported in part by Award No. RP1-187 of the
U.S. Civilian Research \& Development Foundation for the Independent
States of the Former Soviet Union (CRDF) and by Russian Foundation
for Basic Research, grant 96-02-17804a.

}
\section*{Appendix A}
\def\theequation{A\arabic{equation}}
\setcounter{equation}{0}

Here we present the expressions for the functions $W^{(1)}$ 
and $W^{(2)}$ of Eq. (\ref{sigma-W}) in terms of the variables 
$\tau$, $\chi$ and $\gamma$ (see Eq. (\ref{x-tau})). The function 
$W^{(1)}$ reads
\begin{eqnarray}
 & & \frac{1}{\lambda} W^{(1)}(\tau, \rho, \chi, \gamma) =  
 E \chi - N \ln \gamma +  R_{b}^{*} \frac{T}{1 - \gamma X} R_{b} 
                   \label{W1} \\
 & + & \gamma R_{a}^{*} \frac{X T^{-1}}{1 - \gamma X} R_{a} + 
 \gamma R_{a} \frac{X}{1 - \gamma X} R_{b} +
 \gamma R_{a}^{*} \frac{X}{1 - \gamma X} R_{b}^{*}, \nonumber
\end{eqnarray}
where $R_{a}$ and $R_{b}$ are the residues of the instanton field on
the mass shell corresponding to initial and final particles,
respectively (in the notations of Ref. \cite{Ti2}). Here 
\[
X(\bp, \bk) = \delta(\bp - \bk) e^{-\omega_p\chi}, \; \; \; \;
T(\bp, \bk) = \delta(\bp - \bk) e^{-\omega_p\tau}, 
\]
and integration 
\[
    \int \frac{d\bp \ d\bk}{\sqrt{\omega_{\bp}\omega_{\bk}}}
\]
is assumed where appropriate. For our model $R_{a}(\bp)=R_{b}(\bp)$ 
and equals the expression (\ref{R-inst}) with 
$\omega_{\bp}=\sqrt{\bp^{2} + m^{2}}$, so that (\ref{W1}) 
can be written as 
\begin{eqnarray}
& & \frac{1}{\lambda} W^{(1)}(\tau, \rho, \chi, \gamma) =  
 E \chi - N \ln \gamma \label{W1-1}  \nonumber \\
 & + & \frac{192 \pi^{2} \rho^{2}}{\lambda} \left[ 
 J(\gamma, \tau, \chi) + \gamma J(\gamma, \chi-\tau, \chi) + 
 2 \gamma J(\gamma, \chi, \chi) \right], 
\end{eqnarray}
where 
\[
  J(\gamma, \tau, \chi) = \frac{1}{4 \pi} 
  \int \frac{d \bk} {\omega_{\bk}} 
  \frac{e^{-\omega_{\bk}\tau}}{1 - \gamma e^{-\omega_{\bk}\chi}}. 
\] 

The next-to-leading order function $W^{(2)}$ can be written as the sum
of contributions representing interactions of final-final,
initial-final and initial-initial particles, respectively,
\begin{equation}
W^{(2)} =  W^{(2)}_{(f-f)} + W^{(2)}_{(i-f)} + W^{(2)}_{(i-i)} ,
\label{W2}
\end{equation}
where 
\begin{eqnarray}
\frac{1}{\lambda} W^{(2)}_{(f-f)} & = &  \frac{1}{2} \biggl[
 R_{b} \frac{T}{1-\gamma X} D_{bb}^{\dagger}\frac{T}{1-\gamma X}R_{b} 
\nonumber\\
& & + R_{b}^{*} \frac{T}{1-\gamma X} D_{bb} \frac{T}{1-\gamma X}R_{b}^{*}
 + \ldots \biggr],  \nonumber \\
\frac{1}{\lambda} W^{(2)}_{(i-f)}& = &  \frac{\gamma}{2} \biggl[
 R_{b} \frac{T}{1-\gamma X} D_{ab}^{\dagger}\frac{X T^{-1}}{1-\gamma X}
    R_{a} \nonumber\\
& &
 + R_{a}^{*} \frac{X T^{-1}}{1-\gamma X} D_{ab} \frac{T}{1-\gamma X}
      R_{b}^{*}  + \ldots \biggr],  \nonumber \\
\frac{1}{\lambda} W^{(2)}_{(i-i)}& = & \frac{\gamma^{2}}{2} \biggl[
 R_{a} \frac{X T^{-1}}{1-\gamma X} D_{aa}^{\dagger}
      \frac{X T^{-1}}{1-\gamma X}R_{a}  \nonumber\\
& &
 + R_{a}^{*} \frac{X T^{-1}}{1-\gamma X} D_{aa} 
 \frac{X T^{-1}}{1-\gamma X}R_{a}^{*}   
 + \ldots \biggl]. \nonumber  
\end{eqnarray} 
Only terms relevant in the small $\nu$ limit are shown explicitly.
$D_{aa}$, $D_{ab}$ and $D_{bb}$ are related to the double
on-mass-shell residue of the propagator, Eq.(\ref{res}), as follows,
\[
D_{\#}(\bp, \bk) = \frac{\rho^{2}} {(2 \pi)^{3} 2 \sqrt{\omega_{\bp}
  \omega_{\bk}}} R(\rho^{2} s_{\#}(\bp, \bk)),
\]
where $\# = aa,\ ab,\ bb$ and $s_{\#}$ is the $s$-variable for 
corresponding particles on the mass shell,
\begin{eqnarray}
s_{aa}(\bp, \bk) & = & s_{bb}(\bp, \bk) = (p + k)^{2} = 
2 m^{2} - 2[\omega_{\bp} \omega_{\bk} - \bp\bk ], \nonumber \\
s_{ab}(\bp, \bk) & = & (p - k)^{2} = 
2 m^{2} + 2[\omega_{\bp} \omega_{\bk} - \bp\bk ]. \nonumber
\end{eqnarray}
One has
\begin{eqnarray}
W^{(2)}_{(f-f)} & = & 384 \pi^{2} (\rho m)^{4} \Bigl[ 
  J_{bb}(\gamma, \tau, \tau, \chi) + 
  2 \gamma J_{bb}(\gamma, \tau, \chi, \chi) + \nonumber\\
& &
  \gamma^{2} J_{bb}(\gamma, \chi, \chi, \chi) \Bigr], 
      \nonumber \\
W^{(2)}_{(i-f)} & = & 768 \pi^{2} (\rho m)^{4} \gamma \Bigl[ 
  J_{ab}(\gamma, \tau, \chi, \chi) + J_{ab}(\gamma,\tau,\chi-\tau,\chi) \nonumber \\
  &+&  \gamma J_{ab}(\gamma, \chi, \chi, \chi) 
   +  
  \gamma J_{ab}(\gamma,\chi-\tau, \chi, \chi) \Bigr], 
      \nonumber \\
W^{(2)}_{(i-i)} & = & 384 \pi^{2} (\rho m)^{4} \gamma^{2} \Bigl[ 
  J_{aa}(\gamma, \chi-\tau, \chi-\tau, \chi) + 
  J_{aa}(\gamma, \chi-\tau, \chi, \chi) \nonumber \\ 
  & + &  
  J_{bb}(\gamma, \chi, \chi, \chi) \Bigr], 
      \nonumber 
\end{eqnarray}
where
\[
J_{\#}(\gamma, \tau_{1}, \tau_{2}, \chi) = 
\frac{1}{8\pi^{2}} \int \frac{d \bk}{\omega_{\bk}} 
\frac{d \bq}{\omega_{\bq}}
\frac{e^{-\omega_{\bk}\tau_{1}}}{1-\gamma e^{-\omega_{\bk}\chi}} 
\frac{ R \left(\rho^{2} s_{\#}(\bk,\bq) \right)}{16 \pi^{2}} 
\frac{e^{-\omega_{\bq}\tau_{2}}}{1-\gamma e^{-\omega_{\bq}\chi}}.
\]
For $m \tau_{1,2} \ll 1$ and $\gamma = 0$ one finds
\begin{eqnarray} 
J_{aa}(0,\tau_{1},\tau_{2},\chi) & = & 
  J_{bb}(0,\tau_{1},\tau_{2},\chi) \nonumber \\ 
  & = & 384 \pi^{2} \rho^{4} 
  \left[ -\frac{3}{\tau_{1}^{2} \tau_{2}^{2}} 
  \left( \frac{58}{15} 
  + \ln \frac{\rho^{2}}{\tau_{1}\tau_{2}} \right) - 
  \frac{12 \rho^{2}}{\tau_{1}^{3} \tau_{2}^{3}} 
  \left( \frac{71}{30} 
  + \ln \frac{\rho^{2}}{\tau_{1}\tau_{2}} \right) \right] ,
  \nonumber \\
  J_{ab}(0,\tau_{1},\tau_{2},\chi) & = & 
  384 \pi^{2} \rho^{4} 
  \left[ -\frac{3}{\tau_{1}^{2} \tau_{2}^{2}} 
  \left( \frac{58}{15} 
  + \ln \frac{\rho^{2}}{\tau_{1}\tau_{2}} \right) + 
  \frac{12 \rho^{2}}{\tau_{1}^{3} \tau_{2}^{3}} 
  \left( \frac{71}{30} 
  + \ln \frac{\rho^{2}}{\tau_{1}\tau_{2}} \right) \right] .
  \nonumber 
\end{eqnarray}

\end{document}